# Constraints on terrestrial planet formation timescales and equilibration processes in the Grand Tack scenario from Hf-W isotopic evolution

Nicholas G. Zube[1,*], Francis Nimmo[1], Rebecca A. Fischer[2], Seth A. Jacobson[3]

1. University of California Santa Cruz, Dept. of Earth and Planetary Sciences, 1156 High St., Santa Cruz, CA 95064, USA (nzube@ucsc.edu, fnimmo@es.ucsc.edu).

2. Harvard University, Dept. Earth and Planetary Sciences, 20 Oxford St., Cambridge, MA 02138, USA (rebeccafischer@g.harvard.edu).

3. Northwestern University, Dept. Earth and Planetary Sciences, 2145 Sheridan Road, Evanston, IL 60208, USA (sethajacobson@earth.northwestern.edu).

**Abstract**

We examine 141 N-body simulations of terrestrial planet late-stage accretion that use the Grand Tack scenario, coupling the collisional results with a hafnium-tungsten (Hf-W) isotopic evolution model. Accretion in the Grand Tack scenario results in faster planet formation than classical accretion models because of higher planetesimal surface density induced by a migrating Jupiter. Planetary embryos that grow rapidly experience radiogenic ingrowth of mantle $^{182}$W that is inconsistent with the measured terrestrial composition, unless much of the tungsten is removed by an impactor core that mixes thoroughly with the target mantle. For physically Earth-like surviving planets, we find that the fraction of equilibrating impactor core $k_{core} \geq 0.6$ is required to produce results agreeing with observed terrestrial tungsten anomalies (assuming equilibration with relatively large volumes of target mantle material; smaller equilibrating mantle volumes would require even larger $k_{core}$). This requirement of substantial core re-equilibration may be difficult to reconcile with fluid dynamical predictions and hydrocode simulations of mixing during large impacts, and hence this result does not favor the rapid planet building that results from Grand Tack accretion.

**1. Introduction**

The hafnium-tungsten (Hf-W) isotopic system has become a standard chronometer for determining timescales of core formation during late-stage planetary accretion (e.g. Jacobsen, 2005; Kleine et al., 2009; Kleine and Walker, 2017). In this time period, the isotopic signature of a terrestrial body evolves during its growth due to accretion of smaller planetesimals and larger embryos (Halliday, 2004; Jacobsen, 2005). For a given Hf/W ratio in the mantle, a higher relative concentration of mantle $^{182}$W results when accretion finishes earlier. Therefore, measurements of this isotopic system for terrestrial bodies provide insight into the chronology of planetary accretion in the solar system (Kleine et al., 2009; Rudge et al, 2010; Morishima et al., 2013; Jacobson et al., 2014).

Interpreting the Hf-W system is complicated by two factors. One is the stochastic, multi-stage growth of planetary bodies. The other, and arguably more serious problem, is the uncertain degree of re-equilibration between core-forming metals and mantle silicates following each impact. Larger degrees of re-equilibration drive down the signature of radiogenic $^{182}$W in the mantle, potentially erasing the high concentration of mantle $^{182}$W expected from a body that experienced rapid early growth. The physical processes determining the extent of re-equilibration at a planetary scale are not yet well understood, and we will discuss the implications of re-equilibration on the Hf-W isotopic system in Section 2.3.

The stochastic growth of embryos to a final planetary state complicates the isotopic evolution of the core and mantle. Timing of impacts, relative masses of impactors and targets, silicate mass fraction, initial states of colliding bodies, and metal-silicate partitioning conditions in the bodies make each major collisional event a contributing factor to their final isotopic states. Analytical models with steady planetary growth fail to capture the punctuated effects of multiple large collisions. Instead, investigation of late-stage accretion can be carried out in numerical simulations of collisional growth (e.g. Agnor et al., 1999; Chambers, 2001; O'Brien et al., 2006; Raymond et al., 2006), allowing constraints like mass, semi-major axis, and eccentricity of extant planets to be used to evaluate such scenarios in their ability to reproduce the arrangement of the inner solar system.

A documented shortcoming found in classical accretion models is the production of Mars analogs with masses that are an order of magnitude too large (Wetherill, 1991; Chambers, 2001; O'Brien et al., 2006; Raymond et al., 2006). Variations on the classical scenario with Jupiter and Saturn on eccentric orbits produce slightly overlarge Mars analogs and dry terrestrial planets with excited orbits (Chambers and Cassen, 2002; Raymond et al., 2009). While the small size of Mars could be a low-probability outcome of classical accretion scenarios (Fischer and Ciesla, 2014), an alternative is the Grand Tack model, in which a gas-driven inward-then-outward migration of Jupiter and Saturn reduces the surface density of planetesimals in the region between Mars and Jupiter (Walsh et al., 2011). Grand Tack simulations have reproduced Mars-sized planets and the compositional structure of the asteroid belt (Hansen, 2009; Walsh et al., 2011; O'Brien et al., 2014). An important feature of the Grand Tack results is that increased planetesimal and embryo surface densities around 1 AU due to Jupiter's migration lead to more collisions and faster planet building than the classical scenarios.

An additional test of the Grand Tack scenario is its consequences for the Hf-W system. In this study, we examine the effect that rapid accretion in the Grand Tack scenario has on Hf-W isotopic evolution in Earth and Mars analogs. This work expands on the Hf-W isotopic calculations used in Nimmo and Agnor (2006) and Nimmo et al. (2010) that examined classical N-body models of late-stage accretion. A full explanation of tungsten isotope systematics and classical N-body scenarios are found in these papers. As we will describe below, the rapid accretion characteristic of Grand Tack simulations results in Earth-like planets that generally develop tungsten anomalies larger than the measured terrestrial value. Unless substantial re-equilibration between impactor cores and the proto-Earth's mantle occurs during collisions, the Grand Tack results are inconsistent with the tungsten isotopic signature of the Earth.

## 2. Methods

*2.1 $^{182}$Hf-$^{182}$W system*

Hafnium is lithophile while W is moderately siderophile, which results in strong fractionation during a core formation event. Unstable $^{182}$Hf decays to $^{182}$W with a half-life of 9 My. An early core formation event will strip most of the tungsten present in the mantle away to the core while hafnium remains. After time passes, this results in a mantle with excess $^{182}$W compared to stable isotopes such as $^{183}$W or $^{184}$W, referred to here as a positive mantle tungsten anomaly. Conversely, a late core formation event could remove most W from the mantle, depending on how thoroughly impactor material re-equilibrates with the target's mantle during the impact. The tungsten isotopic evolution of planetary bodies is thus controlled by the timing of impacts, the degree of tungsten partitioning into a target core, and the degree of core-mantle re-equilibration during collisions. The dependence of final tungsten anomaly on collisional history is complex and stochastic, making a large suite of simulations desirable in determining how changes in the initial conditions will affect the resulting isotopic signatures.

There are two key measurable quantities pertinent to the Hf-W system (e.g. Kleine et al., 2009). We define the ratio of Hf to W in a planetary mantle relative to a chondritic reference as the mantle fractionation factor, $f^{Hf/W}$:

$$f^{Hf/W} = \frac{(C^{180Hf}/C^{183W})_{mantle}}{(C^{180Hf}/C^{183W})_{CHUR}} - 1 \qquad (1)$$

$C^{180Hf}$ and $C^{183W}$ are mantle concentrations of the isotopes, while Chondritic Uniform Reservoir (CHUR) refers to an undifferentiated chondritic ratio. In the interest of comparing our results with previous Hf-W analyses done on classical accretion models, we adopt the isotopic values used in Nimmo et al. (2010). We use the chondritic $^{180}$Hf/$^{183}$W ratio = 2.627, the bulk silicate Earth $f^{Hf/W}$ = 13.6 ± 4.3 (Kleine et al., 2009), and Mars $f^{Hf/W}$ = 2.4 ± 0.9 (Nimmo and Kleine, 2007). The second constraint is the mantle tungsten anomaly $\varepsilon_W$:

$$\varepsilon_W = \left[\frac{(C^{182W}/C^{183W})_{mantle}}{(C^{182W}/C^{183W})_{CHUR}} - 1\right] \times 10^4 \qquad (2)$$

We note that these definitions are different from studies that report W isotopic data relative to a terrestrial standard. Isotopic anomalies arise during the coincident occurrence of fractionation and radioactive decay. When fractionation occurs while the radioactive parent element ($^{182}$Hf) is extant, the subsequent growth of the daughter ($^{182}$W) in the mantle will be detectable at a later time. The measured value for the mantle tungsten anomaly in the present Earth is $\varepsilon_W$ = 1.9 ± 0.1 (Kleine et al., 2009). A late veneer likely added material to the bulk silicate Earth after the core finished forming at the end of late-stage accretion, reducing the tungsten anomaly by 0.27 ± 0.04 epsilon units (Kruijer et al., 2015). For this study, we therefore take the end of late-stage accretion $\varepsilon_W$ value of the Earth mantle to be 2.2 ± 0.15. For Mars, shergottite sources yield a tungsten anomaly of $\varepsilon_W$ = 2.3 ± 0.2 (Kleine et al., 2004; Foley et al., 2005). These values have generous error compared to the most recent studies (e.g. Kruijer and Kleine, 2017), but we note that late veneer calculations are somewhat model dependent. The already difficult prospect of obtaining the Earth's $\varepsilon_W$ calls for a conservative definition of uncertainties since we desire significant numbers for a statistical approach.

*2.2 Grand Tack N-Body accretion model*

We apply tungsten isotope calculations as a post-processing step on results from a suite of 141 Grand Tack N-body simulations performed by Jacobson and Morbidelli (2014). The simulations begin with a protoplanetary disk of about 30 to a few hundred Mars-sized embryos and a few thousand smaller planetesimals at ~2.4 My after the condensation of the first solids in the solar system. At the start of these simulations, inward migration of Jupiter and Saturn occurs during the first 0.1 My, followed by Saturn reaching its final mass and driving outward migration for the next 0.5 My until the gas in the disc is removed and migration halts. The initial mass ratio of embryos to planetesimals is set to be one of 1:1, 2:1, 4:1, or 8:1, and the initial mass of individual embryos is varied as 0.025, 0.05, or 0.08 $M_E$, where $M_E$ represents one Earth mass. These starting embryo sizes are also approximately equivalent to 0.25, 0.5, and 0.8 $M_{Mars}$. Each combination of the two initial mass variables is repeatedly run through 10–18 separate simulations. Orbits and collisions are then tracked for 150 My, generally resulting in ~4 terrestrial planets on stable orbits. Collisions are treated as perfect mergers with no fragmentation for simplicity. Note that classical simulations that include fragmentation typically take longer to complete planetary growth, and thus yield slightly smaller tungsten anomalies (<0.2 $\varepsilon$ units on average) compared to those that do not (Dwyer et al., 2015). Finally, a more complete account of the Grand Tack simulation parameters used here may be found in Jacobson and Morbidelli (2014).

*2.3 Equilibration*

We define the equilibration factor $k_{core}$ ($0 \leq k_{core} \leq 1$) as the fraction of impactor core material that equilibrates with the target mantle prior to entering the target core. In this arrangement, $k_{core} = 1$ indicates complete re-equilibration, while $k_{core} = 0$ indicates no re-equilibration (core merging). Lower equilibration results in less W leaving the mantle during collisions. Here we initially fix $k_{core}$ so that all impacts in a single simulation equilibrate to the same degree, and later relax this assumption. The equilibration process is very poorly understood on the scale of planetesimal and embryo impacts. Physical effects of equilibration and turbulent entrainment occur on a cm-scale (Rubie et al., 2003), which cannot be resolved in planetary-scale hydrodynamic simulations (e.g. Marchi et al., 2018).

In addition to the fraction of the core that re-equilibrates with the mantle, $k_{core}$, the relative volume of mantle material encountered by the sinking core also influences the resulting composition and the final $\varepsilon_W$ (Morishima et al., 2013; Deguen et al., 2014; Fischer and Nimmo, 2018). This second factor is likely to vary based on conditions of the individual collision (Rubie et al., 2003; Rubie et al., 2015). For example, the core of a large impactor may sink without significant fragmentation or turbulent entrainment to merge with a target core, thereby interacting with only a relatively small volume of material in the target mantle. This example's metal dilution factor, which is the ratio of the mass of equilibrated silicate to the mass of equilibrated metal, would be quite small. Laboratory experiments have examined the extent of this kind of re-equilibration through the mechanism of turbulent entrainment, deriving a relationship between core and interacting mantle volumes (Deguen et al., 2011; Deguen et al., 2014).

We examine whether adjusting the equilibration factor based on the relative volume of interacting mantle significantly affects the final mantle tungsten anomaly. In principle, we could modify the isotopic partitioning calculations to take this effect into account directly, as in Rubie et al. (2015) and Fischer and Nimmo (2018). For simplicity, however, we instead follow the approach of Deguen et al. (2014, Equations A.6 and 6) and calculate the influence of metal dilution on equilibration for each impact as follows:

$$\delta_W = 1 + \frac{D_W}{\Delta} \qquad (3)$$

$$\Delta = \frac{\rho_{sil}}{\rho_{met}} \left[ \left(1 + \alpha \frac{z}{r_0}\right)^3 - 1 \right] \qquad (4)$$

Here we refer to $\delta_W$ as the diluted equilibration coefficient, which reduces overall re-equilibration when $\Delta$, the metal dilution factor (mass ratio of equilibrated silicate to equilibrated metal), is sufficiently low. If the metal interacts with only a small volume of silicates, the total mass exchanged, and thus the degree of re-equilibration, is reduced. The physical quantities include: the tungsten partition coefficient $D_W = \frac{C_{core}}{C_{mantle}}$ (see Section 2.4), the ratio of silicate density to metal density $\frac{\rho_{sil}}{\rho_{met}} \sim 0.5$, the entrainment coefficient determined through experiment $\alpha \sim 0.26$ (Deguen et al., 2014), the depth traveled in molten silicate $z$, and the radius of metal blob $r_0$. We assume that an impactor core has a length scale approximately ½ of its body's radius, $r_I$, and that a target mantle has a depth of approximately ½ of its body's radius, $r_T$. Thus, we note that the scale relationship of $\left(\frac{z}{r_0}\right)$ is roughly equivalent to $\left(\frac{r_T}{r_I}\right)$. To convert $\left(\frac{z}{r_0}\right)$ to a mass ratio of impactor:target, $\gamma$, we substitute assuming similar bulk densities for both objects:

$$\gamma = \frac{M_I}{M_T} = \frac{\rho V_I}{\rho V_T} = \frac{4/3 \pi r_I^3}{4/3 \pi r_T^3} = \left(\frac{r_I}{r_T}\right)^3 \sim \left(\frac{z}{r_0}\right)^{-3}$$

and find:

$$\Delta = \frac{\rho_{sil}}{\rho_{met}}\left[\left(1 + \frac{\alpha}{\gamma^{1/3}}\right)^3 - 1\right] \tag{5}$$

The physical picture of the diluted equilibration model first considers the interacting volume of impactor core and the interacting volume of target mantle. After the equilibration of these two components, the equilibrated portion of the core sinks to the center of the target and mixes with the target core + non-interacting fraction of the impactor core. Similarly, the equilibrated portion of the mantle mixes with the rest of the impactor mantle + non-interacting target mantle. The experimentally-determined dilution described by Equation (4) may be considered a lower bound, since additional mixing effects could occur in planetary-scale accretionary impacts or in material ponding at the bottom of a magma ocean.

Following Deguen et al. (2014), this process is represented by the full equilibration factor, $k$:

$$k = k_{core}/\delta_W \tag{6}$$

This term is referred to by Deguen et al. (2014) as the equilibration efficiency, $\varepsilon_i$. Note that since for all impacts $1/\delta_W < 1$, it is always the case that $k < k_{core}$. In keeping with earlier work, we will typically specify re-equilibration in terms of $k$ and assume the metal dilution $\Delta$ is infinite, such that $\delta_W = 1$ and thus $k = k_{core}$. Allowing $k$ to vary with impact mass ratio generally has the effect of reducing overall re-equilibration significantly, which is a topic we return to in the Discussion below.

Fig. 1a shows the result of applying Equations (3) and (5) to derive the influence of diluted equilibration on $k$ for different $\gamma$, assuming constant $D_W$. This plot shows that larger

Zube, Nimmo, Fischer, Jacobson    *Hf/W in Grand Tack (2018)*                                                                      11

impactors equilibrate with relatively less mantle material, and thus experience a smaller *k*. In reality, $D_W$ likely decreases over the course of accretion (Wade and Wood, 2005; Fischer and Nimmo, 2018) and thus *k* tends to increase as accretion proceeds, other things being equal.

To understand the effect Equation (3) will have on re-equilibration during accretion, it is necessary to characterize the frequency and size of impacts experienced by Earth-like bodies. Fig. 1b shows counts of $\gamma$ for impacts experienced by Earth-like bodies in our Grand Tack simulations. The counts are first weighted by the ratio of impactor mass to final body mass, showing the influence of the impact on the overall material in the body, and then normalized to the total mass of all Earth-like bodies in the simulations, such that the sum of all counts is 1. Over 70% of accreted mass is delivered by impacts in the $10^{-2} \leq \gamma \leq 1$ range, for which *k* will be close to zero (Fig. 1a). Thus, the effect of including $\delta_W$ that varies with impact size will be an overall reduction of re-equilibration occurring in Earth-like bodies.

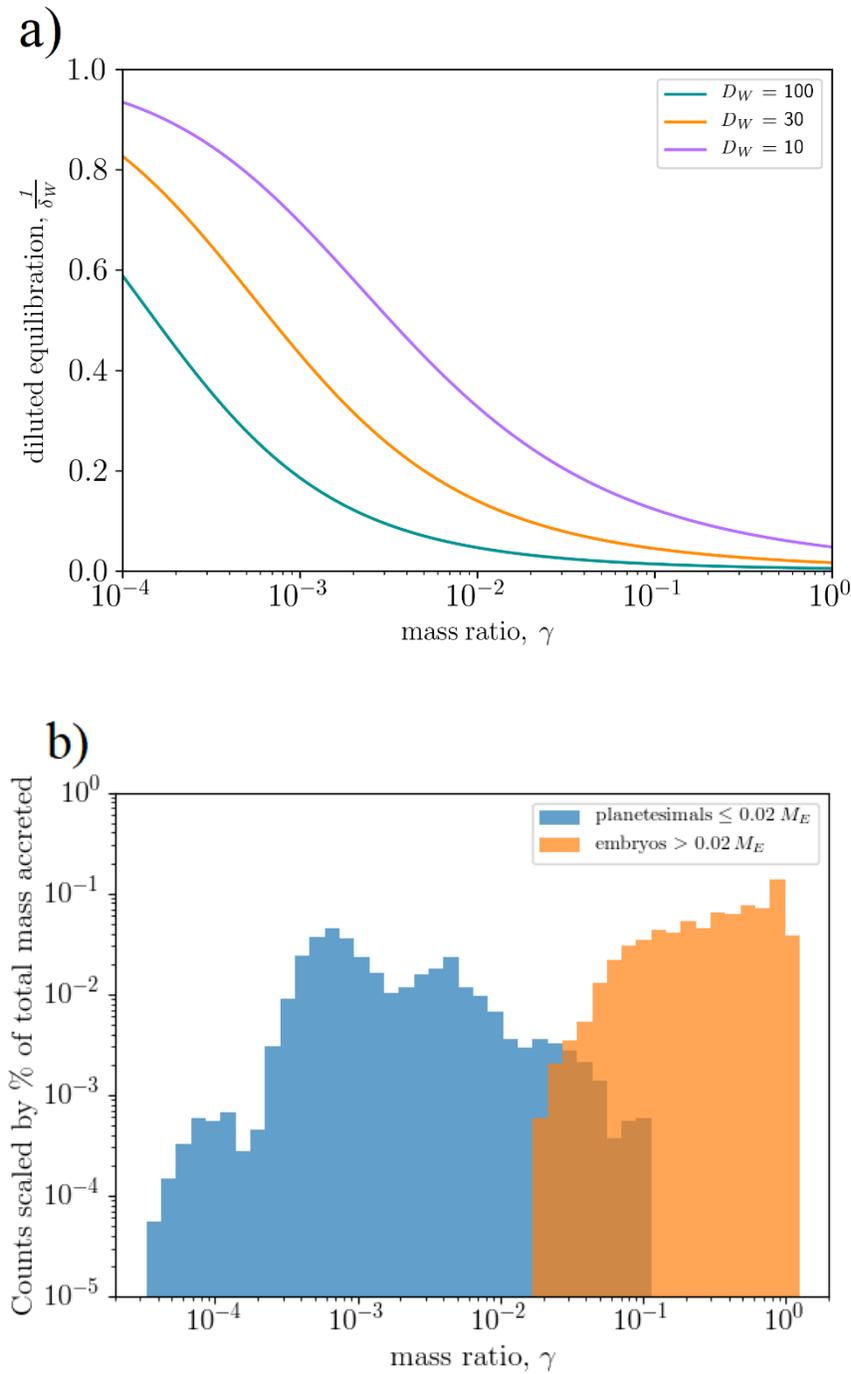

Fig. 1. a) Diluted equilibration $1/\delta_W$ calculated as a function of impactor:target mass ratio using Equations (3) and (5), shown over the range of typical model conditions for impacts. The total amount of equilibration is influenced by dilution according to Equation (6): $k = k_{core} / \delta_W$. For reference, $\gamma = 10^{-4}$ corresponds to a Ceres-sized impactor and an Earth-sized target. b) Counts of

impactor:target mass ratio for all impacts on bodies with a final Earth-like mass, weighted by ratio of impactor mass to final body mass and normalized to the total mass accreted for all Earth-like bodies. Impacts with $\gamma \geq 10^{-2}$ make up ~70% of all accreted mass.

*2.4 Partitioning*

We define the partition coefficient $D_i$ of element $i$ in relation to concentration $C^i$:

$$D_i = \frac{C^i_{core}}{C^i_{mantle}} \quad (7)$$

This coefficient captures the fraction of atoms of a given element that will partition into the equilibrating metal or silicate during core differentiation. After differentiation, the mantle fractionation factor $f^{Hf/W}$ (Equation (1)) is directly related to the partition coefficient by $D_W = f^{Hf/W}\left(\frac{y}{1-y}\right)$ if $D_W$ is assumed to be constant, where $y$ is the silicate mass fraction (Jacobsen, 2005; Equation 49). Though partitioning of W into the core can depend on factors like oxygen fugacity, temperature, and pressure, we instead simulate variations in $D_W$ by allowing the initial $f^{Hf/W}$ to vary based on radial distance from the sun. This approach allows us to generate both Earth-like and Mars-like $f^{Hf/W}$ values. Further justification for this approach is discussed in Section 2.5 and Nimmo et al. (2010). Note that studies of classical accretion scenarios that have varied $D_W$ based on physical parameters in individual impacts produce nearly identical results to those that held it constant (Nimmo et al., 2010; Fischer and Nimmo, 2018). For all collisions in this study, $D_W = 25.51$. Regarding Hf, we take $D_{Hf}$ to be $10^{-4}$ (highly lithophile). We assume all bodies have the same initial silicate mass fraction ($y = 0.68$), which is appropriate for terrestrial planets. This value may vary for actual planetesimals, but Nimmo and Agnor (2006) found that allowing the likely variability in $y$ did not affect the final isotopic results.

*2.5 Initial Hf/W Ratio*

The evolution of four isotopes ($^{182}$Hf and $^{182}$W parent-daughter system, and $^{180}$Hf and $^{183}$W stable isotopes) are tracked for the core and mantle of each body in the N-body simulation using the same initial ratios as in Nimmo et al. (2010). Decay occurs continuously, while each collision modifies the isotopic concentrations for the new body based on an appropriate mixing scenario between the target mantle and the impactor core and mantle. The Grand Tack simulations are defined to begin 0.6 My before the dissipation of the gas disk, which is assumed to occur 3 My after the appearance of solids (calcium-aluminum-rich inclusions, CAIs) in the solar system. The N-body simulations thus have a start time of 2.4 My after CAIs. We allow $^{182}$Hf to decay during this 2.4 My before collisions begin. Bodies are assumed to differentiate during the first impact (e.g. Nimmo and Agnor, 2006); allowing differentiation to happen earlier (i.e. prior to 2.4 My after CAIs) would result in larger final tungsten anomalies (specifically, ~0.1 epsilon units larger for $k = 0.6$ in classical scenarios for differentiation at CAI formation; Fischer and Nimmo, 2018). Following Nimmo et al. (2010), we also assume the initial mantle fractionation factor $f^{Hf/W}$ varies with radial position:

$$f^{Hf/W}{}_i = f_0 + C * \tan^{-1}\left(\frac{1.3 - a}{0.3}\right) \qquad (8)$$

where $f_0 = 13$ is the value at 1 AU, and $a$ is initial semi-major axis (AU). The constant $C$ follows a step function, with $C = 24/\pi$ for $a \geq 1.3$ AU. Inside of 1.3 AU, $C$ varies with $k_{core}$, taking on values of 9.9, 6.5, 5.7, 4.2, 3.4, and 2.7 for $k_{core} = 0.0, 0.2, 0.4, 0.6, 0.8,$ and 1.0, respectively. These values were chosen so that the mean $f^{Hf/W}$ of resulting Earths are approximately equal to the observed value of 13.6 ± 4.3, and resulting Mars-like bodies near 1.5 AU are approximately equal to the observed Mars value of 2.4 ± 0.9 (Kleine et al., 2004;

Jacobsen, 2005; Nimmo and Kleine, 2007). Equation (8) is ad hoc and is primarily intended to reproduce the range of $f^{Hf/W}$ values measured in the inner solar system. Interior to 1.3 AU, initial $f^{Hf/W}$ is scaled down with increasing $k_{core}$ so that the average $f^{Hf/W}$ of surviving Earth-like bodies approximately matches the terrestrial value. Such a scaling is not required beyond 1.3 AU to produce Mars-like bodies with $f^{Hf/W}$ near the Mars value, so the shape of the function in that region is constant over all simulations.

## 3. Results

The physical and isotopic results of the 141 N-body simulations are shown in Tables 1 and 2, respectively, for Earth-like bodies of mass $0.5 \leq M_E \leq 2.0$ and semi-major axis between the present-day orbits of Mercury and Mars (0.387 AU and 1.524 AU). Time is measured from after the formation of CAIs in the solar system (Section 2.2) so that the model starts at 2.4 My. We also require Mars-like bodies to have $0.5 \leq M_{Mars} \leq 2.0$ and semi-major axes of 1.0–2.3 AU. As a reminder, these models assume $\delta_W = 1$ (metal dilution is infinite) and $k = k_{core}$, in keeping with past studies.

Table 1: Summary of average physical results for surviving Earth-like planets. $M_i$ is the initial embryo mass (in $M_E$), $M_f$ is the final body mass ($M_E$), $a_f$ is the final semi-major axis (AU), $t_{95}$ is the simulation time of accretion of 95% of the final mass (My), and $n$ is the number of Earth-like bodies over all simulations of this type. All variables list the mean and standard deviation with the exception of $t_{95}$, which uses a median and middle 67th percentile because of a few large outliers being present in growth timescales. E:P is the initial mass ratio of embryos to planetesimals.

| E:P | $M_i$ | $M_f$ | $a_f$ | $t_{95}$ | n |
|---|---|---|---|---|---|
| 1:1 | 0.025 | 0.77 ± 0.17 | 0.81 ± 0.17 | 46.5 ± 8.9 | 21 |
| 1:1 | 0.05 | 0.85 ± 0.18 | 0.78 ± 0.14 | 44.6 ± 9.4 | 18 |
| 1:1 | 0.08 | 0.67 ± 0.10 | 0.81 ± 0.15 | 46.9 ± 6.0 | 11 |
| 2:1 | 0.025 | 0.82 ± 0.13 | 0.79 ± 0.16 | 41.3 ± 8.7 | 19 |
| 2:1 | 0.05 | 0.75 ± 0.18 | 0.82 ± 0.19 | 41.1 ± 10.0 | 21 |
| 2:1 | 0.08 | 0.80 ± 0.28 | 0.82 ± 0.14 | 39.5 ± 6.7 | 13 |
| 4:1 | 0.025 | 0.87 ± 0.21 | 0.85 ± 0.22 | 42.6 ± 28.4 | 18 |
| 4:1 | 0.05 | 0.89 ± 0.28 | 0.82 ± 0.21 | 34.7 ± 24.7 | 19 |
| 4:1 | 0.08 | 0.81 ± 0.20 | 0.84 ± 0.19 | 35.3 ± 7.7 | 18 |
| 8:1 | 0.025 | 1.00 ± 0.33 | 0.84 ± 0.22 | 50.4 ± 24.9 | 18 |
| 8:1 | 0.05 | 0.95 ± 0.31 | 0.83 ± 0.20 | 28.9 ± 46.9 | 36 |
| 8:1 | 0.08 | 1.09 ± 0.38 | 0.84 ± 0.21 | 25.5 ± 38.7 | 26 |

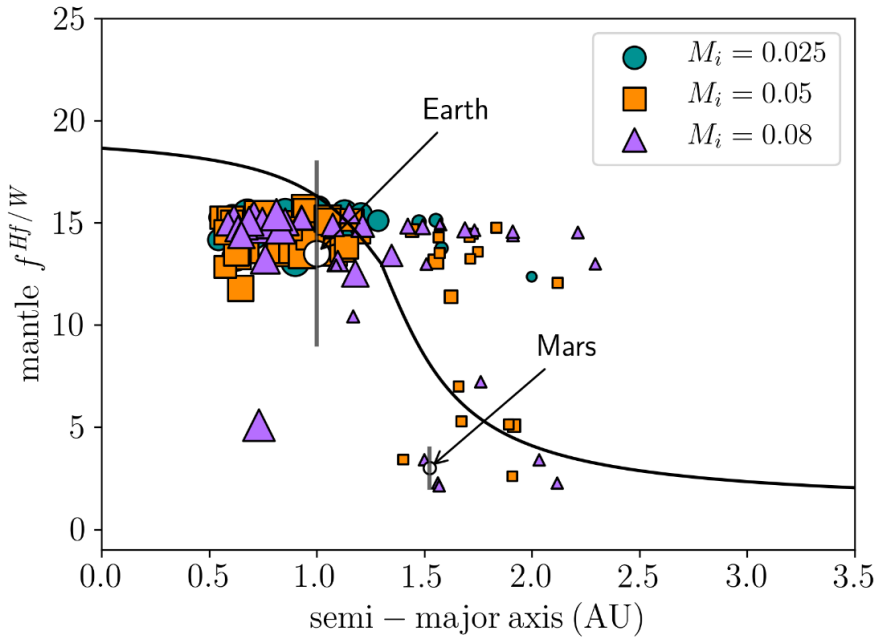

Fig 2. Final mantle $f^{Hf/W}$ against final semi-major axis for surviving planets in $k = 0.6$ simulations with initial 8:1 embryo:planetesimal mass ratio. Symbol size is proportional to final mass, and shape/color denote initial embryo mass $M_i$. Earth and Mars values are shown as white circles with uncertainty in mantle $f^{Hf/W}$ represented by a grey line. Initial $f^{Hf/W}$ distribution is shown as a solid line.

Fig. 2 plots the final $f^{Hf/W}$ against final semi-major axis for surviving planets in simulations with initial 8:1 embryo:planetesimal mass ratio. The solid line denotes the initial $f^{Hf/W}$ curve (Equation (7)). By design, bodies around 1 AU have $f^{Hf/W}$ values comparable to that of the Earth. Bodies around 1.5 AU show more variability in $f^{Hf/W}$ than for equivalent classical accretion scenarios (cf. Fig. 1a of Nimmo et al., 2010), indicating the greater degree of radial mixing that takes place in the Grand Tack scenario (e.g. Fischer et al., 2018). Nonetheless, at least some bodies at Mars-like semi-major axes achieve Mars-like values of $f^{Hf/W}$.

Table 2: Mean and standard deviation of the isotopic results for surviving Earth-like bodies. $M_i$ is initial embryo mass in $M_E$, $\varepsilon_W$ and $f^{Hf/W}$ are the mean values for the final bodies, and the ratios are the initial embryo:planetesimal mass distribution.

|  | $M_i = 0.025$ | | $M_i = 0.05$ | | $M_i = 0.08$ | |
| --- | --- | --- | --- | --- | --- | --- |
|  | $f^{Hf/W}$ | $\varepsilon_W$ | $f^{Hf/W}$ | $\varepsilon_W$ | $f^{Hf/W}$ | $\varepsilon_W$ |
| ($k=1.0$) | | | | | | |
| 1:1 | 14.2 ± 0.2 | 1.5 ± 0.6 | 14.2 ± 0.2 | 1.7 ± 0.6 | 14.0 ± 0.3 | 1.5 ± 0.4 |
| 2:1 | 13.7 ± 2.1 | 1.9 ± 0.5 | 14.1 ± 0.4 | 2.1 ± 0.9 | 13.9 ± 0.5 | 2.2 ± 0.7 |
| 4:1 | 14.1 ± 0.3 | 2.0 ± 0.8 | 13.6 ± 2.4 | 2.5 ± 0.9 | 14.1 ± 0.4 | 2.9 ± 1.0 |
| 8:1 | 14.1 ± 0.5 | 2.0 ± 0.9 | 13.9 ± 0.6 | 3.2 ± 1.8 | 13.8 ± 1.9 | 3.2 ± 1.7 |
| ($k=0.8$) | | | | | | |
| 1:1 | 14.6 ± 0.3 | 1.9 ± 0.7 | 14.6 ± 0.2 | 2.1 ± 0.7 | 14.3 ± 0.4 | 1.9 ± 0.4 |
| 2:1 | 14.0 ± 2.2 | 2.3 ± 0.6 | 14.4 ± 0.6 | 2.6 ± 1.1 | 14.2 ± 0.6 | 2.8 ± 0.9 |
| 4:1 | 14.5 ± 0.4 | 2.5 ± 1.0 | 14.0 ± 2.4 | 3.1 ± 1.1 | 14.3 ± 0.5 | 3.6 ± 1.2 |
| 8:1 | 14.5 ± 0.6 | 2.4 ± 1.0 | 14.2 ± 0.7 | 3.7 ± 2.0 | 14.1 ± 1.9 | 3.8 ± 1.9 |
| ($k=0.6$) | | | | | | |
| 1:1 | 14.9 ± 0.3 | 2.6 ± 0.8 | 14.8 ± 0.3 | 2.8 ± 0.9 | 14.4 ± 0.5 | 2.6 ± 0.5 |
| 2:1 | 14.2 ± 2.2 | 3.0 ± 0.7 | 14.6 ± 0.7 | 3.4 ± 1.3 | 14.3 ± 0.7 | 3.6 ± 1.1 |
| 4:1 | 14.9 ± 0.5 | 3.1 ± 1.2 | 14.2 ± 2.5 | 3.9 ± 1.3 | 14.4 ± 0.6 | 4.5 ± 1.3 |
| 8:1 | 14.8 ± 0.8 | 3.1 ± 1.2 | 14.4 ± 0.9 | 4.5 ± 2.2 | 14.3 ± 2.0 | 4.7 ± 2.1 |
| ($k=0.4$) | | | | | | |
| 1:1 | 15.4 ± 0.5 | 3.6 ± 1.0 | 15.3 ± 0.4 | 4.0 ± 1.2 | 14.7 ± 0.6 | 3.8 ± 0.7 |
| 2:1 | 14.7 ± 2.2 | 4.1 ± 0.8 | 15.0 ± 0.9 | 4.7 ± 1.6 | 14.6 ± 0.9 | 4.9 ± 1.3 |
| 4:1 | 15.5 ± 0.6 | 4.3 ± 1.4 | 14.7 ± 2.6 | 5.2 ± 1.6 | 14.6 ± 0.8 | 6.0 ± 1.5 |
| 8:1 | 15.4 ± 1.0 | 4.3 ± 1.4 | 14.8 ± 1.2 | 5.8 ± 2.4 | 14.9 ± 2.1 | 6.0 ± 2.3 |

A key characteristic of the Grand Tack simulations is that planets finish growing more rapidly than in classical simulations. In general, Earth-like bodies have finished building 95% of their mass by 30–40 My in the Grand Tack, whereas classical simulations finish after (median) 64 My with a large spread (Nimmo et al., 2010). Other things being equal, if planetary growth is completed more rapidly, a larger mantle tungsten anomaly should be the result. The groups with generally shorter 95% growth timescales (e.g. 4:1 and 8:1 with 0.05 and 0.08 $M_i$; Table 1) produce Earths with the highest average $\varepsilon_W$, as expected (Table 2). Because typical planetary growth curves change with both initial embryo:planetesimal ratio and initial embryo mass, defining a growth timescale is not trivial. We do not use the timing of the last giant impact because several bodies experience significant development in Hf/W isotope evolution after a final embryo-embryo collision, either from $^{182}$Hf remaining in the mantle or continued planetesimal-embryo collisions. The timescale $t_{95}$ was chosen since most bodies experience little Hf/W evolution beyond this point, and it is very rare for any embryo-embryo collisions to occur after this time.

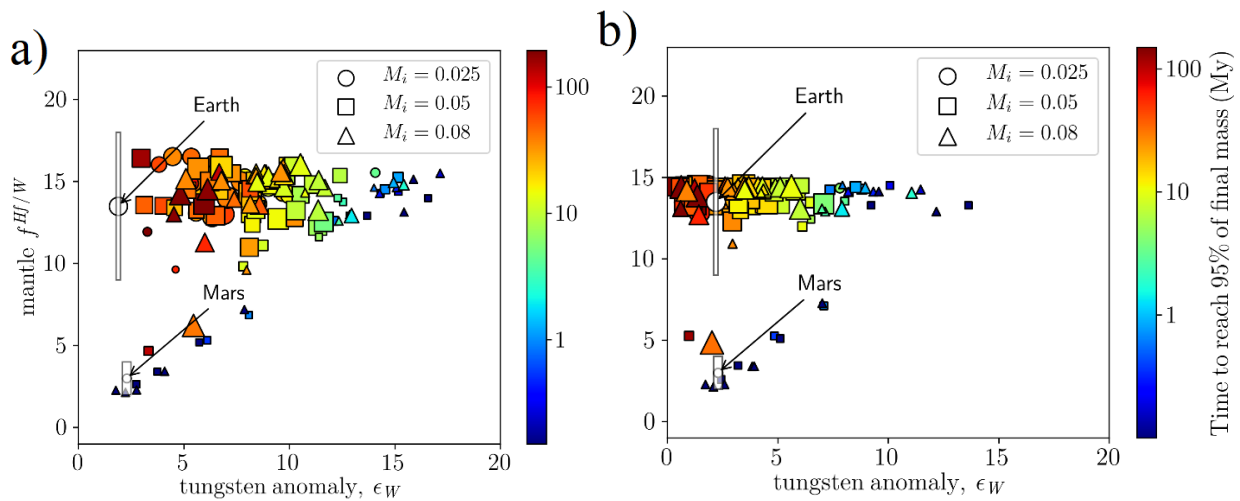

Fig. 3. a) Variation of mantle tungsten anomaly $\varepsilon_W$ with respect to mantle $f^{Hf/W}$, using equilibration factor $k = 0.2$ (very little re-equilibration) and starting embryo:planetesimal mass ratio 8:1. Symbol size is proportional to final mass, and shape denotes initial embryo mass $M_i$. Color denotes the time when a body attains 95% of its final mass. Earth and Mars values are shown as white circles with uncertainty represented by white rectangles. b) Same as a), but with equilibration factor $k = 1.0$ (complete re-equilibration).

Fig. 3 plots the variation in mantle $f^{Hf/W}$ against the mantle tungsten anomaly $\varepsilon_W$, with Fig. 3a and Fig. 3b using equilibration factor $k = 0.2$ and $k = 1.0$, respectively. We remind the reader that these models assume complete silicate equilibration ($\delta_W = 1$). In general, only small bodies exhibit low $f^{Hf/W}$ values. A body that grows significantly beyond its initial size will evolve based on the characteristics of the material being accreted, and there is a larger quantity of material available from the inner part of the disk where $f^{Hf/W}$ is large by assumption (Fig. 2). If a body only experiences a single-stage core formation, there is a linear correlation between $f^{Hf/W}$ and $\varepsilon_W$, because in this case the mantle tungsten anomaly simply depends on the $^{182}Hf/^{180}Hf$ ratio in the mantle at the time of core formation (e.g. Jacobsen, 2005). However, Fig. 3 shows no such correlation except for bodies that experience very little growth. For larger bodies, high $f^{Hf/W}$ does not necessarily result in high $\varepsilon_W$ because these bodies experience more collisions. During impacts with high enough equilibration $k$, mixing will drive down the tungsten anomaly in the mantle. Fig. 3b shows that the Grand Tack produces several Earth-like analogs with approximately correct $\varepsilon_W$ under conditions of complete re-equilibration ($k = 1.0$). Bodies that take the longest to reach 95% of final mass are the most strongly influenced by the $\varepsilon_W$ reduction due to re-equilibration, as expected. Conversely, Fig. 3a shows results from low equilibration ($k = 0.2$), yielding much higher tungsten anomalies. In this case, even the rare statistical outliers that finish building after 100 My do not produce an isotopically correct terrestrial mantle.

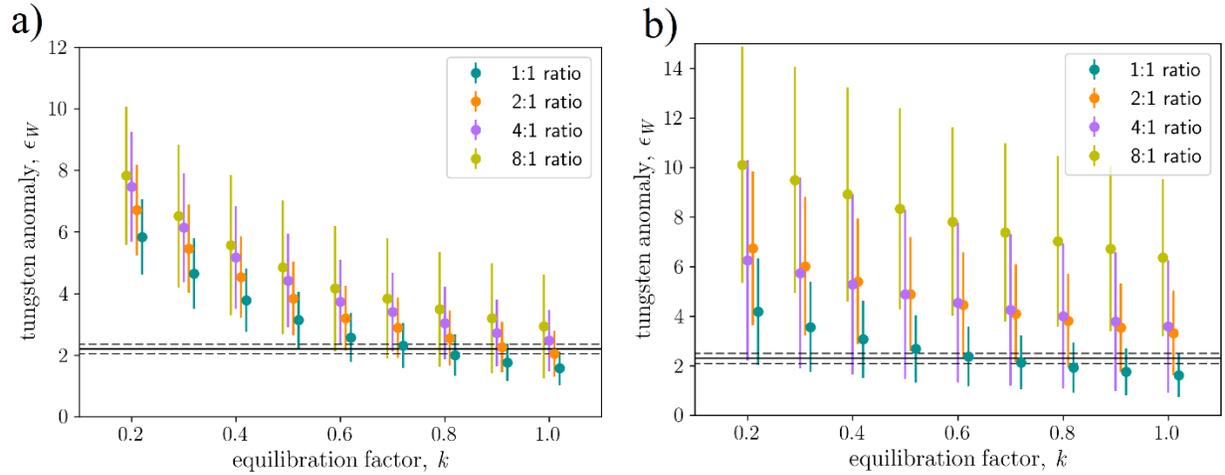

Fig. 4. a) Mean and standard deviation of mantle tungsten anomaly $\varepsilon_W$ as a function of equilibration factor $k$, calculated using all Earth-like bodies with parameters defined in Table 2. Color represents the initial embryo:planetesimal mass ratio, and each group is horizontally offset for clarity. Solid and dashed lines represent the pre-late veneer value for the Earth and its uncertainty, respectively (see Section 2.1). As a reminder, these models assume $\delta_W = 1$ (metal dilution is infinite) and $k = k_{core}$, in keeping with past studies. b) Same as a), but for Mars-like bodies, as defined at the beginning of Section 3, and solid and dashed lines representing the measured Mars value and uncertainty defined in Section 2.1.

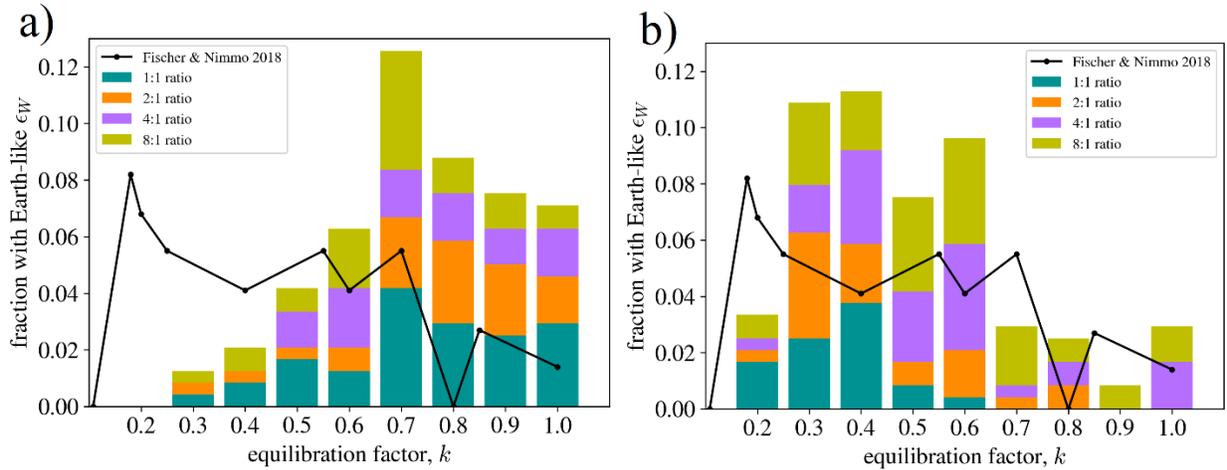

Fig 5. a) Fraction of total Earth-like bodies that match the pre-late veneer terrestrial tungsten anomaly ($\varepsilon_W = 2.2 \pm 0.15$). The solid line shows results from classical accretion scenarios in Fischer and Nimmo (2018). b) Like a), but with simulated classical accretion timescales where each collision occurs twice as late as when collisions occurred in the Grand Tack simulations.

The effect of varying $k$ on the average tungsten anomaly for Earth-like bodies is shown in Fig. 4a. The Grand Tack models require that $k \geq 0.6$ (substantial re-equilibration) to produce bodies with Earth-like $\varepsilon_W$ on average. The most successful reproduction of terrestrial $\varepsilon_W$ occurs when $k$ is near 0.7-0.8 (depending on initial mass ratio), which is significantly higher than classical models where the ideal isotopic outcome was produced with $k \sim 0.4$ (Nimmo and Agnor, 2006; Nimmo et al., 2010; Rudge et al., 2010; Fischer and Nimmo, 2018). Groups of simulations with higher initial embryo:planetesimal mass ratios also have Earth-like bodies that generally complete 95% growth in a shorter period (Table 1), which is reflected in the consistently higher average $\varepsilon_W$ for these simulations (Table 2).

The simulations also produce smaller, Mars-mass bodies with a wide range of values for $\varepsilon_W$. It is worth noting that in simulations where embryos begin with a mass of 0.5 or 0.8 $M_{Mars}$, such bodies experience collisions with only planetesimals since any embryo-embryo collision would exceed the limits of the final allowed Mars mass. These models are thus unlikely to reflect the actual tungsten evolution that Mars experienced during late accretion growth. However, cases with initial embryo masses of 0.25 $M_{Mars}$ can experience embryo-embryo collisions during the simulation, and thus perhaps have a growth history more comparable to that of the actual Mars. Fig. 4b shows that simulated $\varepsilon_W$ values are generally higher than values for the real Mars; however, we did not find any systematic influence of $k$ on the probability of obtaining a Mars-like tungsten anomaly. This latter result is probably because of the wide range of $f^{Hf/W}$ values that arise for Mars-like objects in our approach (Fig. 2).

An alternative way of assessing the ability of Grand Tack-style models to reproduce the Earth is shown in Fig 5a. This plots the fraction of Earth-like bodies that match the terrestrial tungsten anomaly of $\varepsilon_W = 2.2 \pm 0.15$. The advantage of this approach is that it allows a conditional (Bayesian) question to be posed: given that the Earth has the tungsten anomaly it does, what is the expected value of $k$? Fig 5a. shows that even though all realizations result in a rather low (<13%) fraction of Earth-like values, the most likely value of $k$ is 0.7. This result may be compared with equivalent results from classical accretion scenarios, which are shown as the solid line in Fig. 5a. On the whole, both types of scenarios are fairly unlikely to produce Earth-like $\varepsilon_W$ regardless of the degree of re-equilibration, but classical accretion succeeds in more than 5% of cases for the range $0.2 < k < 0.7$, while the Grand Tack scenario is most likely to succeed for $0.6 < k < 1.0$. We remind the reader that these models assume infinite metal dilution ($k = k_{core}$).

The Grand Tack requires high degrees of re-equilibration because of the rapid growth of Earth analogs. To test the effect of accretion time on final mantle tungsten anomaly, we also performed the same Hf/W analysis on Grand Tack simulations with all collisions occurring at $2t_{col}$ instead of $t_{col}$ (where $t_{col}$ is the collision time from the original simulations). These "double time" growth timescales for Earth analogs approximately match the growth timescales of classical accretion. As shown in Fig. 5b, the range of $k$ producing the most bodies with Earth-like $\varepsilon_W$ in "double time" simulations becomes 0.3–0.6, which agrees with the results of Fischer and Nimmo (2018) for classical simulations.

It is possible that the nebular gas disk could have persisted to up to 5 My after CAIs (Johnson et al., 2016). Changing the assumed N-body simulation start time from 2.4 My to 4.4 My after CAIs has a predictable effect on the final tungsten anomalies of Earth-like bodies (this timing maintains 0.6 My before gas disk dissipation). There is less $^{182}$Hf remaining when collisions begin, and overall tungsten anomalies are slightly (10–20%) lower. However, this does not significantly change the number of Earth-like bodies matching the terrestrial $\varepsilon_W$ value. Plotting the frequency of Earth-like $\varepsilon_W$ results in a very similar picture to Fig. 4a: simulations using $k < 0.6$ produce Earth-like anomalies in less than 5% of cases, and the likelihood of producing Earth-like bodies when $k \geq 0.6$ remains in the 8–12% range. We also tested the idea that lower initial $f^{Hf/W}$ could result in reduced $\varepsilon_W$ for surviving bodies. Setting the amplitude of the initial $f^{Hf/W}$ distribution to the lower bound of the terrestrial value within uncertainty (9.3) produces slightly lower values of final $\varepsilon_W$, but not enough to change the range of acceptable equilibration factors ($k \geq 0.6$). This adjustment caused almost no change in the likelihood of producing a body with Earth-like $\varepsilon_W$ shown in Fig. 5a. However, it became much more unlikely to produce a body with Earth-like $f^{Hf/W}$. Invoking lower initial $f^{Hf/W}$ is not an effective mechanism for producing Earth-like isotopic outcomes.

Rather than solely specifying $\delta_W = 1$, we also carried out simulations in which $\delta_W$ was defined using Equation (5). In this approach, the influence of impactor size on $k$ sharply reduces the ability of large impacts to re-equilibrate with the mantle. As shown in Fig. 1a, impacts with $\gamma$ (impactor:target mass ratio) larger than 0.2 and $D_W = 30$ will result in $k \leq 0.2$. Since these impacts with $\gamma \geq 0.2$ make up ~70% of all accreted mass (Fig. 1b), the effective result of including a varying $\delta_W$ is to cause a lower overall equilibration factor. In these simulations, the final $\varepsilon_W$ follows a generally linear relationship with $f^{Hf/W}$, as expected for single-stage core formation. Even for the ideal case of $k_{core} = 1$ (which produces the greatest possible value of $k$), it is effectively impossible to lower $\varepsilon_W$ to the measured terrestrial value in Grand Tack simulations if silicate equilibration in planetary impacts is correctly described by Equation (4). Only if large impacts result in significant metal and silicate equilibration can the measured terrestrial $\varepsilon_W$ value be reconciled with these Grand Tack accretion simulations.

## 4. Discussion

It is clear from previous results that the degree of re-equilibration is a critical parameter in determining mantle tungsten anomaly (Nimmo and Agnor, 2006; Rudge et al., 2010; Rubie et al., 2015). This re-equilibration depends on both the core mass that re-equilibrates and the mass of target silicate involved (Section 2.3). Both effects can be combined into a comprehensive re-equilibration factor, $k$ (Equation (6)), and below we discuss our results in terms of this combined parameter. The distinction between $k_{core}$ and $k$ only matters for studies (e.g. Rubie et al., 2015; Fischer and Nimmo, 2018) in which re-equilibration is calculated using the ratio of equilibrating silicates to equilibrating metal, rather than assuming infinite dilution ($k = k_{core}$).

Figs. 4 and 5 are the key results of this study: a high equilibration factor ($k \sim 0.7$) is necessary for Grand Tack N-body simulations to be likely to reproduce Earth-like mantle tungsten anomalies. This result can be understood in terms of planetary growth timescales. Other things being equal, more rapid planet growth will always result in a larger tungsten anomaly. Since the Grand Tack builds planets significantly faster than classical accretion scenarios, an additional factor must be acting to drive the tungsten anomaly down. In our approach, this factor is the high degree of re-equilibration. Our models with $k < 0.5$ were not capable of producing any Earth-like $\varepsilon_W$ values with a frequency higher than 4%. In models with $k > 0.6$, this frequency generally raises to 8–12%.

The importance of rapid accretion is further supported by the "double time" Grand Tack simulations that produce results very similar to a model analyzing classical scenarios (Fischer and Nimmo, 2018) in Fig. 5b. Although the two models have notable differences (e.g. fixed vs. variable $f^{Hf/W}$), the fact that they produce such similar outcomes supports the idea that the rapid accretion associated with the Grand Tack causes the requirement of a high degree of re-equilibration.

Past studies examining classical accretion scenarios determined that a median value of $0.3 \leq k \leq 0.8$ was capable of reproducing Earth-like tungsten isotopic characteristics (Nimmo and Agnor, 2006; Nimmo et al., 2010; Rudge et al., 2010). A study that allowed the partition coefficient $D_W$ to evolve with oxygen fugacity and pressure/temperature conditions during collisions still arrived at a similar range (0.2–0.55) of permissible $k$ values (Fischer and Nimmo, 2018). Comparisons of classical cases with fixed (Nimmo et al., 2010) and variable (Fischer and Nimmo, 2018) $f^{Hf/W}$ resulted in only very minor differences. This is because the models are constrained to reach a final $f^{Hf/W}$ value equal to the measured terrestrial value. The collisional history of bodies and the effect of equilibration during mixing therefore play a much larger role in the isotopic outcome.

Rubie et al. (2015) used Grand Tack accretion simulations and found that a high degree of metal re-equilibration ($k_{core} > 0.7$) was required to match mantle concentrations of moderately siderophile elements. However, they also assumed a silicate re-equilibration fraction governed by Equation (5) and thus their overall $k$ was smaller (Equation (6)). Fig. 1 suggests that, assuming $D_W = 30$ and $\gamma > 10^{-3}$, $k$ was in the range ~0.3–0.5. This is approximately consistent with constraints on $k$ derived by Fischer and Nimmo (2018) using classical accretion simulations and matching the mantle W concentration. Only radiogenic elements (such as tungsten) are sensitive to the timescale of accretion; the low $k$ values obtained by Rubie et al. (2015) thus suggest that their results will—like this study—also have difficulty generating low enough tungsten anomalies for their model Earths. However, Fischer et al. (2017) found that higher degrees of both metal and silicate equilibration (larger $k$) can still be consistent with the siderophile element composition of Earth's mantle, due to strong tradeoffs between the degree and depth of equilibration.

Re-equilibration at the planetary scale remains poorly understood. Fluid dynamics experiments show that little equilibration occurs when a metal body is comparable in size to the length of its path though silicate material (Deguen et al., 2014). If these results are applicable to planetary collisions, this precludes the possibility of near-complete re-equilibration during impacts. Further, fluid dynamical arguments presented by Dahl and Stevenson (2010) suggest that 20% or less of each core should equilibrate with silicate material during impacts, regardless of impact velocity. Hydrodynamic simulations of collisions where the target is less than one order of magnitude larger than the impactor result in most of the impactor core material passing rapidly to the target core (Marchi et al., 2018), which is equivalent to low *k*. Even in cases where the target was much larger, glancing impacts (that resulted in impactor core fracture or slow, circumferential paths through the mantle) were required for significant mixing of impactor core and target mantle to occur (Kendall and Melosh, 2012). However, these simulations are not able to reach a centimeter-scale resolution where equilibration and turbulent mixing physically occur (Rubie et al., 2007). As of now, it is not clear physically how a high degree of equilibration—required by Grand Tack simulations—could occur during planetesimal and embryo collisions. Certainly, the results presented in Fig. 1 suggest that reaching $k \geq 0.6$, as required by the tungsten anomalies, is very difficult to achieve.

Such equilibration would, in contrast, be a natural outcome of "pebble accretion" scenarios, in which planets are formed out of dm-scale objects (e.g. Johansen et al., 2015). However, pebble accretion results in very rapid growth, which could lead to an even higher tungsten anomaly. Furthermore, in the absence of giant impacts, features of terrestrial planets such as the tilt of the Earth, the existence of the Moon, and the significantly reduced mantle mass of Mercury become hard to understand. More research is clearly necessary, but until a convincing physical argument for near-total re-equilibration is found, the necessity of high *k* presents a potential problem for the Grand Tack accretion scenario.

## 5. Conclusion

The Grand Tack N-body simulations analyzed here can produce bodies with tungsten isotopic characteristics close to Earth and Mars, but in the case of Earth require the condition of near-complete core-mantle re-equilibration ($k \sim 0.7$). This arises because the Earth analogs complete their growth within around 20–50 My, more quickly than in classical accretion scenarios. To reduce mantle $\varepsilon_W$ to terrestrial levels, moderate to high equilibration is necessary.

Both classical and Grand Tack scenarios are limited to producing isotopically correct Earths with a probability of ~10% even at optimal equilibration values. The main distinction between the models' ability to achieve correct $\varepsilon_W$ is that classical scenarios produce Earths with correct tungsten anomalies most frequently at lower $k$ (0.2–0.6), while Grand Tack scenarios require higher $k$ (0.6–1.0) to achieve a better than 5% rate of isotopically correct Earths. This contrast is driven by the difference in time it takes for planets to complete growth in the two scenarios.

When the influence of non-infinite metal dilution is considered ($\delta_W$ as defined by Equation (5)), which has the general effect of reducing $k$, it becomes impossible to generate terrestrial $\varepsilon_W$ values in the Grand Tack scenario. Even assuming total core re-equilibration ($k_{core} = 1$) and $D_W = 30$, $\Delta > 45$ is necessary for $k$ to achieve a value of 0.6 or higher (Fig. 1a). Using Equation (5), this $\Delta$ is equivalent to most mass being delivered to Earth with an impactor:target mass ratio of $4 \times 10^{-4}$, which is unlikely given the frequency of embryo-embryo impacts that occur in classical and Grand Tack simulations (Fig. 1b). A mechanism for high equilibration has not yet been demonstrated in fluid dynamical experiments (Deguen et al., 2014) or hydrodynamic collision simulations (Marchi et al., 2018), though it is possible that collisions at a very high incidence angle could increase an impactor's path through the target mantle enough that effective equilibration would be enhanced (Kendall and Melosh, 2012). In any case, it is improbable for the Grand Tack to produce Earth-like bodies with terrestrial $\varepsilon_W$ values without relying on a high degree of re-equilibration in all impacts. An alternative model—which has the dynamical characteristics of the Grand Tack without producing rapid terrestrial planet accretion—would be desirable.

**Acknowledgements:**

We would like to thank Thorsten Kleine and an anonymous reviewer for helpful comments. This work was supported by NASA Headquarters under the NASA Earth and Space Science Fellowship Program – Grant 80NSSC17K0479 and by Grant NNX17AE27G.